\documentclass[aip,pof,showpacs,superscriptaddress,reprint]{revtex4-1}

\usepackage{amssymb,amsmath,amsfonts,graphicx}
\usepackage{bm}
\usepackage{comment}
\usepackage{subfigure}
\usepackage{xcolor}

\definecolor{red}{rgb}{1,0,0}
\definecolor{blue}{rgb}{0,0,1}

\newcommand{\beq}{\begin{equation}}
\newcommand{\eeq}{  \end{equation}}
\newcommand{\beqa}{\begin{eqnarray}}
\newcommand{\eeqa}{  \end{eqnarray}}

\def \be {\begin{equation}}
\def \ee {\end{equation}}
\def \ba {\begin{array}}
\def \ea {\end{array}}
\def \bea {\begin{eqnarray}}
\def \eea {\end{eqnarray}}
\def \beas {\begin{eqnarray*}}
\def \eeas {\end{eqnarray*}}
\def \bsea{\begin{eqnarray}} 
\def \esea{\end{eqnarray}} 

\def \bco {\begin{comment}}
\def \eco {\end{comment}}

\newcommand{\Kappa}{\mathrm{K}}

\newcommand{\tmop}[1]{\ensuremath{\operatorname{#1}}}

\definecolor{grey}{rgb}{0.75,0.75,0.75}
\definecolor{dark green}{rgb}{0.75,0.75,0.75}
\definecolor{orange}{rgb}{1.0,0.5,0.5}
\definecolor{brown}{rgb}{0.5,0.25,0.0}
\definecolor{pink}{rgb}{1.0,0.5,0.5}
\newcommand{\fracd}[2]{\displaystyle\frac{#1}{#2}}

\def\Oh{\tmop{Oh}}

\def \om{\omega} 
 
\def\ha{\hat{a}}

\def\XXint#1#2#3{{\setbox0=\hbox{$#1{#2#3}{\int}$}
     \vcenter{\hbox{$#2#3$}}\kern-.5\wd0}}

\begin{document}

\title{Self-similar impulsive capillary waves on a ligament}

\author{L.~Duchemin$^1$, S. Le Diz\`es$^1$, L. Vincent$^1$ \& E. Villermaux}
\affiliation{Aix Marseille Universit\'e, CNRS, Centrale Marseille, IRPHE UMR 7342, F-13384, Marseille, France}
\affiliation{Institut Universitaire de France}

\date{\today}
  
\begin{abstract}
We study the short-time dynamics of a liquid ligament, held between two solid cylinders, when one is impulsively accelerated along its axis. 
A set of one-dimensional equations in the slender-slope approximation is used to describe the dynamics, including surface tension and viscous effects. 
An exact self-similar solution to the linearized equations is successfully compared to experiments made with millimetric ligaments. 
Another non-linear self-similar solution of the full set of equations is found numerically. Both the linear and non-linear solutions show that the axial depth at which the liquid is affected by the motion of the cylinder scales like $\sqrt{t}$. The non-linear solution presents the peculiar feature that there exists a maximum driving velocity $U^\star$ above which the solution disappears, a phenomenon probably related to the de-pinning of the contact line observed in experiments for large pulling velocities. 
\end{abstract}

%\pacs{47.15.Gf, 47.55.Bx, 68.03.-g, 68.08.-p, 68.15.+e}
\maketitle

%%%%%%%%%%%%%%%%%%%%%%%%%%%%%%%%%%%%%%%%%%%%%%%%%%%%%

When the surface of a cohesive material is moved impulsively, for instance because it is impacted or pulled by another, the core of the material develops internal stresses and deformations. If the material is incompressible like liquids when moderately shocked, it is its boundary itself which must deform. For a solicitation localized in space or time (or both), raises the natural question of the extent and dynamics of the surface perturbation. Classical examples with solids include the celebrated Hertz contact problem, and the Saint-Venant solution for a bar impacted longitudinally, of which Hopkinson gave a variant when it is pulled (see Love \cite{Love} and Graff \cite{Graff1991} for historical references and a detailed exposition). In this precise problem, the stress is transported along the bar at the velocity of sound, according to a  non-dispersive wave equation so that the deformation at any location in the bar reflects that at the impact coordinate an instant earlier.

Waves, however, propagate at the surface of liquids at different speeds according to their wavelength. Kelvin and Rayleigh \cite{Rayleigh_83}, studying the pattern of standing waves at the surface of running water past a fixed obstacle have shown that their wavelength is short, capillary dominated upstream of the obstacle, and that longer wavelengths, gravity dominated, are found downstream of it: dispersive waves are sorted depending on the direction they propagate. 

While the stress penetration into a solid bar allows solving for its stability, and ultimate breakup\cite{Gladd05}, the dispersive dynamics of capillary waves on a liquid ligament helps understanding the remnant liquid mass attached to a rod quickly removed from a pool\cite{Vincent2014}. Impacts distort liquid surfaces into a variety of non-trivial shapes \cite{Cooker1995,Zeff2000,Antkowiak2007}, in particular when they are mediated by capillary waves\cite{VM2007,Vincent2014b}; Here, we provide an original solution for the short time response of a liquid ligament of which one extremity is either pulled, or pushed, and in all cases suddenly accelerated axially.

\begin{figure}[h!]
\begin{center}
	\includegraphics[width=.9 \linewidth]{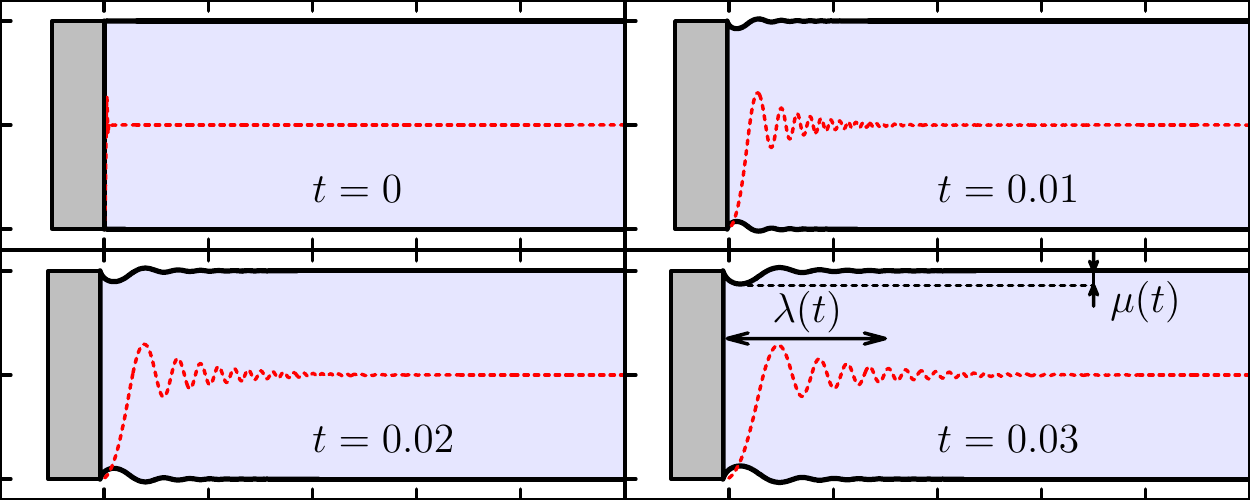}
\end{center}
\caption{Short-time dynamics of the stretched ligament for Oh=$10^{-2}$ and $U_0=-2$. The red dashed curves correspond to the velocity in the ligament. 
The graduations are the same in both directions and the unit distance corresponds to one cylinder radius. 
$\lambda(t)$ is the characteristic depth at which the velocity has decreased significantly 
and is clearly much larger than the cylinder displacement.
}
\label{fig.lig}
\end{figure}

We consider a liquid ligament, initially cylindrical of length $\ell_0$ and radius $R$, held between two cylinders of same radius. The initial state is then characterized by the aspect ratio $\ell_0/R$ and the  Ohnesorge number $\Oh=\eta/\sqrt{\rho \sigma R}$ where $\rho$ is the density of the fluid,  $\sigma$ its surface tension and $\eta$  its dynamic viscosity. One of the  cylinder is impulsively accelerated at $t=0$ along its axis with a velocity $V$, either positive or negative (see figure \ref{fig.lig}).  The strength of this impulsion is measured by the dimensionless parameter $U_0 = V/  \sqrt{\sigma/\rho R}$. In the following,  length and time scales are made dimensionless using  the radius $R$, and the capillary timescale $\tau_c = \sqrt{\rho R^3/\sigma}$. 

We describe the dynamics of the ligament by a one-dimensional model 
\bea 
  (h^2)_t & = & - (h^2 u)_z, \label{equ:NSA} \\
  u_t & = & -u u_z + 
  \Kappa_z + 
  3 \tmop{Oh} \frac{(h^2 u_z)_z}{h^2}, \label{equ:NSu} 
  \eea
  where 
  \be
  \Kappa  =  \frac{h_{zz}}{(1+h_z^2)^{3/2}} - \frac{1}{h(1+h_z^2)^{1/2}},
  \label{equ:kappa}
\ee
is the full curvature of the axisymmetric interface, $u(z,t)$ is the local axial velocity, $h(z,t)$ is the local radius, $z$ is the axial coordinate, and $t$ is the time variable. Although these equations have been derived within the slender-slope approximation\cite{Weber1931,EggersDupont1994}, they have proven to be remarkably efficient at describing the jet dynamics when the slope is far from being small\cite{EggersVillermaux2008}, and even close to the ligament breakup, where the slope diverges\cite{BrennerEggers1997}. Also, choosing the full curvature (\ref{equ:kappa}) instead of the asymptotical limit for small slopes guarantees that the static equilibrium shapes can be correctly recovered. 
This one-dimensional idealization is expected to be relevant to describe the flow in the vicinity of the moving wall at the contact of the rod in $z=0$ where the velocity is, by cohesion, enforced to be uniform, equal to the rod velocity. 
Indeed, we assume in the present analysis that the contact line is fixed to the boundary, an hypothesis justified for small values of $U_0$, but which fails for larger values, as we will see. 

Linearizing equations (\ref{equ:NSA}) and (\ref{equ:NSu}) for small velocities $U_0$ and small deformations around $h_0=1$ gives for the velocity field
\be
2 u_{tt} + u_{zz} + u_{zzzz} - 6 \Oh u_{zzt} = 0,
\label{equ:u}
\ee
from which one deduces the dispersion relation\cite{Weber1931}
\be
-2\om^2 - k^2 + k^4 - 6 i \Oh k^2 \om   =0,
\label{dispersion}
\ee
which connects the frequency $\om$ to the wavenumber $k$ of  a normal mode $e^{i k z - i \om t}$.
The radius $h$ is related to $u$ through $2 h_t=-u_z$. As we expect the dynamics in the first instants to be localized close to the moving boundary, we assume that it is not influenced by the finite size $\ell_0/R$ of the ligament. More precisely, we consider that the other boundary is sufficiently far so that we can make  $\ell_0/R \rightarrow \infty$ in the analysis, an hypothesis on which we come back below.  

It is convenient to solve the problem in the frame moving with the boundary which
experiences an acceleration $U_0 \delta(t)$, {\em i.e.} a velocity jump $U_0 {\rm \mathcal H}(t)$, 
where $\delta(t)$ and ${\rm \mathcal H}(t)$ are the Dirac delta function and the Heaviside step function respectively. 
The acceleration  $a=\partial_t u$ in this {\it non-galilean} reference frame can then be  written as  
\be
a(z,t)= a_A(z,t) - U_0\delta(t) = \frac{U_0}{2\pi}\int_{-\infty}^{\infty} \ha(\om,z) e^{-i\om t} d\om - U_0\delta(t),
\ee
where the boundary conditions on the absolute acceleration $a_A(z,t)$ give for $\ha$
\be 
\ha(\om,0) =1~;~~ \ha(\om,\infty) =0~;~~ \partial_z\ha(\om,0)=0~;~~ \partial_z\ha(\om,\infty)=0.
\label{cond:ha}
\ee
Note that the conditions on $\partial_z \ha$ comes from the fixed contact line condition which requires $\partial_z u$ to vanish on the boundary.
Since $a_A$ satisfies (\ref{equ:u}), we obtain  $\ha$  satisfying (\ref{cond:ha}) as 
\be
\ha (\om,z) = c_+(\om)e^{ik_+(\om)z} + c_-(\om)e^{ik_-(\om)z}
\label{exp:ha}
\ee
with 
\be
c_+ = -\frac{k_-}{k_+-k_-} ~; ~~ c_-= \frac{k_+}{k_+-k_-} ,
\ee
where $k_{\pm}(\om)$ are the two solutions of (\ref{dispersion}) such that $\Im m(k_{\pm}) > 0 $
\be
k_{\pm}(\om) = \sqrt{\frac{ 1 + 6 i \Oh \om \pm \sqrt{ (1 + 6 i \Oh \om )^2 + 8 \om ^2}}{2}} ~.
\ee

The functions $k_{\pm}$ can be analytically defined on the integration contour if it is a line parallel to the $\om_r$ axis above the 
maximum growth rate, that is above, say, $\om_i=1/2$. This choice of contour also guarantees causality: for negative times, the integration contour is 
closed in the upper plane such that the solution is null for $t<0$. From the expression of $a$, we deduce the expressions of $u$ and $h$
\be 
u(z,t) = \frac{U_0}{2\pi} \int_{-\infty}^{\infty} \frac{ c_+e^{ik_+z} + c_-e^{ik_-z}}{-i\om}e^{-i\om t} d\om
 - U_0\,{\rm \mathcal H}(t),
\ee 
\be 
h(z,t) = 1+\frac{U_0}{4\pi} \int_{-\infty}^{\infty} \frac{ c_+k_+e^{ik_+z} + c_-k_-e^{ik_-z}}{-i\om^2}e^{-i\om t} d\om.
\ee
For small times, we expect the solution to vary close to the origin only. In this region, the  
main contribution to the integrals is then expected to come from large values of $|\om|$ for which
$|k_{\pm}(\om)|$ is also large and
given by
\be
k_{\pm} \sim \lambda _{\pm} \sqrt{i\om}  ~~{\rm with} ~~ \lambda _{\pm}^2 = 3 \Oh \pm \sqrt{ 9 \Oh^2 - 2} .
 \ee  
 The functions $\lambda_{\pm}$ are both real and positive for $\Oh > \sqrt{2}/3$, or complex conjugate with  real$(\lambda_{\pm}) >0$ for $\Oh < \sqrt{2}/3$. 
 The coefficients $c_+$ and $c_-$ become independent of $\om$. 
The above expression for $a$, $u$ and $h$ can then be evaluated for $t>0$  using known contour integrals, and we obtain
\be
a(z,t) =  \frac{U_0 z}{2\sqrt{\pi}t\sqrt{t}}\left[c_+\lambda_+ \,e^{-\lambda_+^2 z^2/4t} + 
c_-\lambda_- \,e^{-\lambda_-^2 z^2/4t} \right],
\ee
\be
u(z,t) = -U_0 \left[c_+{\rm Erf}\left(\frac{ \lambda_+ z}{2 \sqrt{t}}\right) + c_-{\rm Erf}\left(\frac{ \lambda_- z}{2 \sqrt{t}}\right)\right] , \\
\ee
\bsea
\ba{ll}
h(z,t)= & 1+U_0 \sqrt{t}  \left[    \fracd{c_+\lambda_+}{\sqrt{\pi}} e^{-\fracd{ \lambda_+^2 z^2}{4t }} - \fracd{ c_+\lambda_+^2 z}{2\sqrt{t}}{\rm Erfc}\left(\fracd{ \lambda_+ z}{2 \sqrt{t}}\right)   \right. \\
& \left. + \fracd{c_-\lambda_-}{\sqrt{\pi}} e^{-\fracd{ \lambda_-^2 z^2}{4t }} - \fracd{c_-\lambda_-^2 z}{2\sqrt{t}}{\rm Erfc}\left(\frac{ \lambda_- z}{2 \sqrt{t}}\right)  \right],
\label{equ:h}
\ea
\esea
Note that  the above expressions are the sum of complex conjugate terms  when $\Oh < \sqrt{2}/3$, while all the terms are  real when $\Oh > \sqrt{2}/3$. 
As soon as $\Oh$ is nonzero, $u$ and $h$ tend exponentially fast to a constant on a length scale proportional to $\sqrt{t/\Oh }$.  For small time, the dynamics is therefore 
limited to the close neighborhood of the moving boundary,  as assumed. More precisely, the finite size of the ligament is not
expected to affect the dynamics if $t$ remains smaller than $\Oh (\ell_0/R)^2$.
\begin{figure}
\begin{center}
	\includegraphics[width=.9 \linewidth]{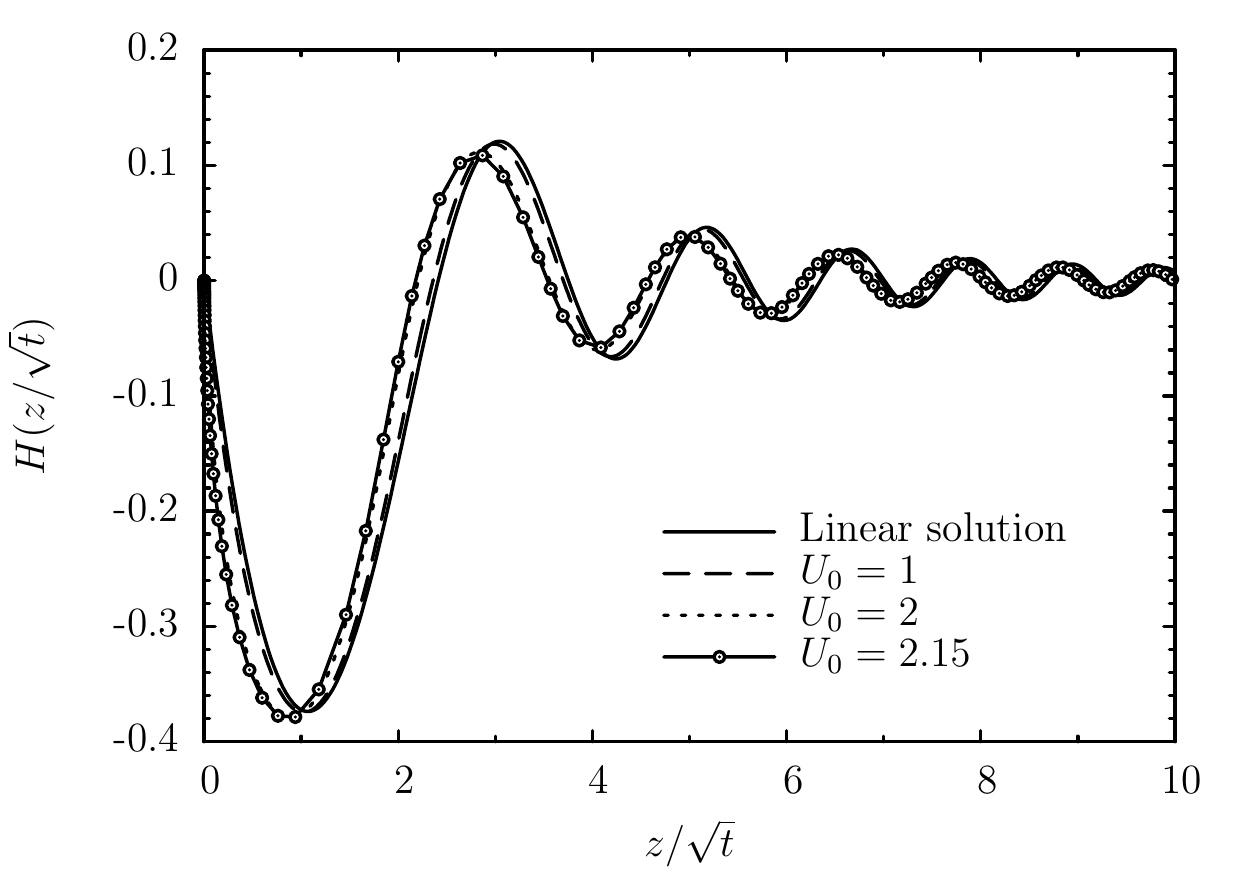}
\end{center}
\caption{Linear and non-linear self-similar solutions for $\Oh = 0.0025$. 
Since $U_0^2$ appears in equations (\ref{selfsim1}) and (\ref{selfsim2}), 
elongated and compressed shapes are symmetric with respect to the unperturbed interface.}
\label{fig.sol}
\end{figure}

It is  worth mentioning that expression (\ref{equ:h}) for $h$ satisfies volume conservation: The ligament volume remains constant up to $O(U_0^2)$ terms since $\delta V= \pi \int_0^\infty (h^2(z)-1) dz \sim \pi U_0 t$  balances the  volume $-\pi U_0 t$ associated with the elongation ($U_0< 0$) or compression ($U_0>0$) of the ligament. Expression (\ref{equ:h}) for $h$ can be written in the self-similar form
\be
h(z,t) = 1 + U_0 \sqrt{t} \times H(z/\sqrt{t}).
\label{equ:ss}
\ee
The scaling variable $z/\sqrt{t}$ reflects the form of the dispersion relation \eqref{dispersion} which is tangent to $\omega^2\sim k^4$ at short time and small scale, a feature shared, in particular, with the dynamics of flexural waves on solid rods\cite{Sneddon,Graff1991,Audoly2005}; the result in \eqref{equ:h} is besides reminiscent of those established by Boussinesq for vibrating elastic beams\cite{Sneddon}. It is genuine to the imposed transverse length scale $R$, and differs from the one known to prevail in surface-tension-driven flows $\omega^2\sim k^3$ in absence of externally imposed length scale\cite{KellerMiksis1983}. Yet, the scaling law in $\sqrt{t}$ can be obtained using the simple ansatz derived from the boundary conditions
\be
u(z,t) = U_0 \, U\left(\frac{z}{\lambda}\right), \; h(z,t)=1 + \mu H\left(\frac{z}{\lambda}\right),
\label{equ:ansatz}
\ee
where $\lambda(t)$ and   $\mu(t)$ are the length scale defined in figure \ref{fig.lig}. 
This ansatz assumes that velocity and surface distorsions follow the same scaling, an hypothesis which is common\cite{Zeff2000}, but not 
systematic \cite{VM2007}. The amplitude scaling for the velocity directly comes from the boundary condition which prescribes that 
$u(z,t)$ should be equal to $-U_0$  at infinity for all time. Plugging (\ref{equ:ansatz}) in equations (\ref{equ:NSA}) and (\ref{equ:NSu}) and requiring the balance of all linear terms, we get
\be
\mu(t) \lambda'(t) \sim \mu'(t) \lambda(t) \sim U_0 \quad \textrm{and} \quad \mu(t) \sim \lambda(t) U_0,
\ee 
giving
\be
\lambda(t) = \sqrt{t} \quad \textrm{and} \quad \mu(t) = U_0 \sqrt{t}.
\label{exp:lambda}
\ee 
\begin{figure}
\begin{center}
	\includegraphics[width=.9 \linewidth]{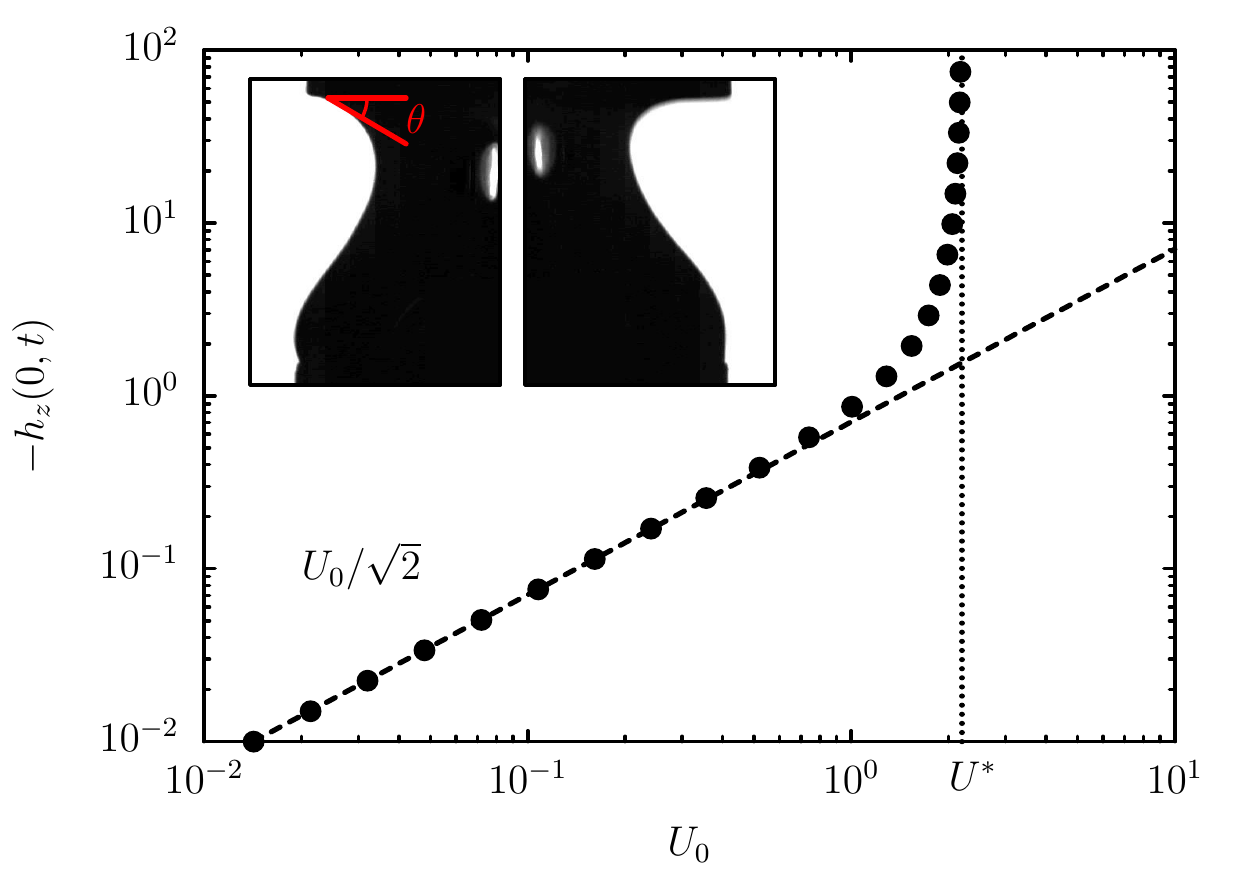}
\end{center}
\caption{Slope of the interface at the contact line ($-h_z(0,t)={\rm cot an}(\theta)$) as a function of the driving velocity $U_0$, for $\tmop{Oh}=10^{-2}$. 
The limiting velocity is $U^\star \simeq 2.2$.
The pictures in inset correspond to experiments for which the driving velocity is smaller than $U^\star$ (left: $U_0\simeq 1.17$) 
and bigger than $U^\star$ (right: $U_0\simeq 3.73$)~: in the later case, the contact line is not pinned anymore 
and dewetting occurs. The red dashed angle drawn on the first figure corresponds to the theoretical slope.} 
\label{fig.fpu}
\end{figure}

\begin{figure}
\begin{center}
	\includegraphics[width=.9\linewidth]{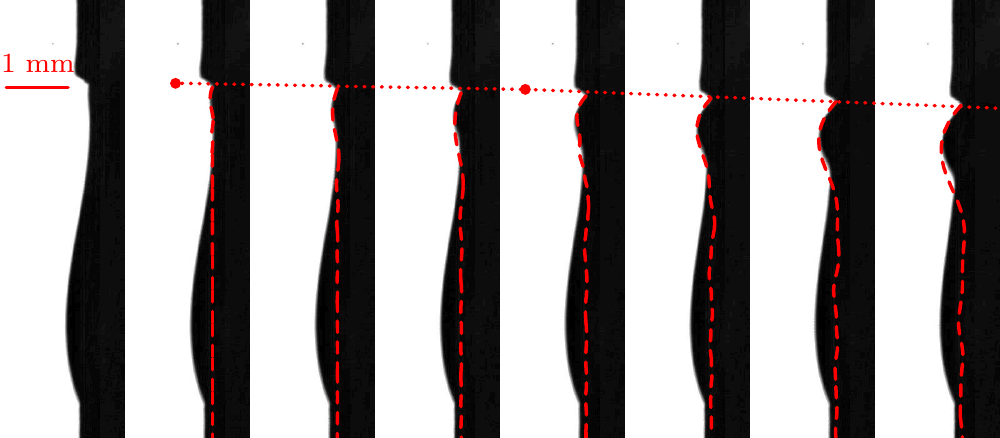}
	\includegraphics[width=.9\linewidth]{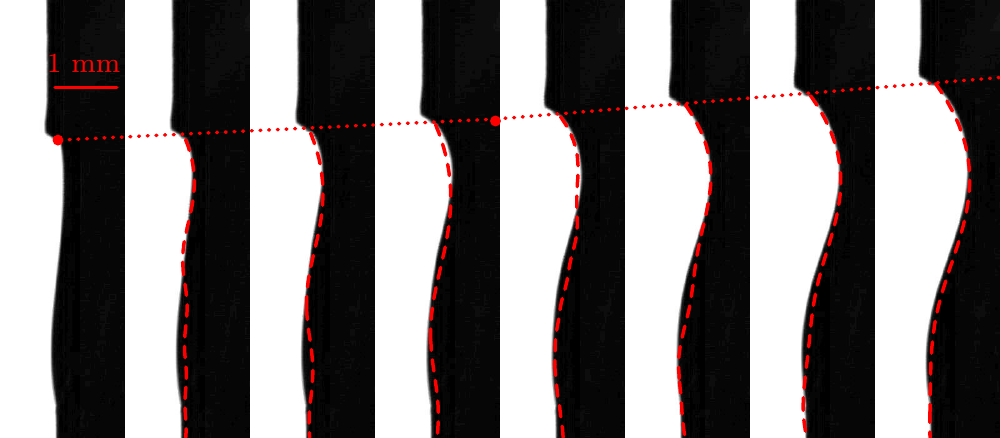}
\end{center}
\caption{Experimental observations of a compressed and a elongated liquid ligament, 
compared to the self-similar solution (\ref{equ:h}) obtained for $\Oh = 0.0025$. 
The radius and capillary time of the experiment are $R=2.5$ mm and $\tau_c =  0.016$ s, respectively.  
{\bf First row~:}  The time interval between two successive frames is $0.2$ ms, that is $\delta t = 1/8$. 
The self-similar solution (red dashed curves) has been obtained as the superimposition of two impulsive solutions, 
the impulses being located at the two red dots. The first red dot ($t_0$) corresponds to a velocity $U_0 = 0.645$
% $V \simeq 0.1$ m/s, 
and the second dot  ($t_1$) to a velocity increment $U_0= 0.821$.
%$V \simeq 0.23$ m/s. 
{\bf Second row~:} The time interval is $0.8$ ms, that is $\delta t =0.5 $ and the two dots correspond 
to velocity increments  $U_0= -0.684$ and $-0.489$ respectively. 
%$V \simeq 0.11$ and $0.18$ m/s respectively.
}
\label{fig.exp}
\end{figure}
Expression (\ref{equ:h}) for $h$ has been obtained in the limit $t \to 0$ and $|U_0| \ll 1$. A more general self-similar 
solution of equations (\ref{equ:NSA}) and (\ref{equ:NSu}) can be obtained by removing the constraint $|U_0| \ll 1$. 
Indeed, inserting  the ansatz (\ref{equ:ansatz}) with (\ref{exp:lambda}) in equations (\ref{equ:NSA}) and (\ref{equ:NSu}), we obtain,  for small times, a system of non-linear equations with respect to the self-similar variable $x=z/\sqrt{t}$ for the function $U$ and $H$
\bea
H(x) - x H'(x) & = & -U'(x),\label{selfsim1}\\
-\frac12 x U'(x) & = & \frac{H'''(x)}{(1+U_0^2 H'^2(x))^{3/2}} 
- \frac{3 U_0^2 H''^2(x) H'(x)}{(1+U_0^2 H'^2(x))^{5/2}}
+ 3 \tmop{Oh} U''(x).
\label{selfsim2}
\eea 
Note that the small $t$ hypothesis guarantees that nonlinearity only appears in the curvature term. This system of ordinary differential equations can be solved numerically. The result of the integration is displayed in figure \ref{fig.sol} for different values of $U_0$ and $\tmop{Oh} =0.0025$.  The nonlinear solution is very close to the linear solution as soon as $U_0 \leq 1 $, a property that we have also observed for other values of $\Oh$.
%%%%%%%%%%%%%%%%%%%%%%%%%%%%%%%%%%%%%%%%%%%%%%%%%%%%%

An interesting feature of equations (\ref{selfsim1}) and (\ref{selfsim2}) arises when seeking a solution for large values 
of the driving velocity $U_0$. Indeed, there exists a limit value of $U_0$ above which there is no solution to this set of equations. 
Figure \ref{fig.fpu} shows the slope of the interface at the contact line, as a function of the driving velocity $U_0$ for $\tmop{Oh} =0.01$. 
For $U_0 \le 1$, the linear behavior described by equation (\ref{equ:h}) is in agreement with the numerical 
results~: the slope of the interface is well approximated by $U_0/\sqrt{2}$, which 
is the value of $-h_z(0,t)$ given by equation (\ref{equ:h}) for $\tmop{Oh} = 0$.

Above $U_0=1$, the solution departs from the linear solution, and finally diverges for a finite value of the velocity 
$U^\star\simeq 2.2$. This limit value corresponds to an interface tangent to the solid surface at the contact line above which the dynamics cannot be described by a self-similar solution anymore. It is not clear what the dynamics would be with a pinned contact line since direct numerical simulations of the initial set of equations (\ref{equ:NSA}) and (\ref{equ:NSu}) systematically lead to a finite time singularity at the boundary. We suspect that this singular behavior is reminiscent of the physical change observed in the experiments. 
When $U_0$ is larger that 1.5, we observe that the contact line is not pinned anymore and starts to move on the solid, as illustrated in the inset of figure \ref{fig.fpu}. 
In this case, the correct formulation would require a model for the moving contact line, relating the contact angle and receding speed of the contact line.

%%%%%%%%%%%%%%%%%%%%%%%%%%%%%%%%%%%%%%%%%%%%%%%%%%%%%

When $U_0$ is smaller than 1.5, the contact line remains attached and the experimental profiles can be compared to the self-similar solution. 
Such a comparison is shown in figure \ref{fig.exp} for a compressed and a elongated configuration.
The experiments is done with a water ligament, hold between two vertical metal (brass) cylinders of diameter $5$ mm. Its initial shape is slightly deformed by gravity.
The upper cylinder is suddenly set into vertical motion and the velocity $U_0$ is obtained by post-processing the experimental movies. 
For both cases, we find that the evolution of the velocity of the upper cylinder can be approximatively reproduced 
by superimposing two different impulses at $t_0$ and $t_1$,  indicated by the red dots in figure \ref{fig.exp}.  
The theoretical solution is then obtained by adding after $t_1$ a second linear solution to the first solution generated at $t_0$. 
In both cases, this linear solution agrees quantitatively with the experimental profiles and the agreement is sufficiently good to confirm the relevance of the self-similar solution. These results also demonstrate that the short time response is barely affected by finite length effects and nonlinear effects.  
This constitutes an a-posteriori validation of the hypotheses made in the theory.

The disappearance of the self-similar solution with a pinned contact line above a critical velocity is an interesting phenomenon
which is worth emphasizing. In the theory, the self-similar solution disappears when the tangent to the boundary becomes vertical, that is the contact angle
vanishes. In experiments, we expect such a  solution to disappear for a smaller velocity, since the contact angle cannot drop below the dynamical contact angle of the receding contact line. For the system water/brass/air that we have used, the dynamical contact angle is approximatively 30 degree, which gives using figure \ref{fig.fpu} a critical velocity $U_c \approx 1.5$, a value consistent with the experimental observations. This, to close, underlines the importance of the boundary conditions in the detailed shape of the distorted ligament surface, although its space-time scaling remains independent of them.

%%%%%%%%%%%%%%%%%%%%%%%%%%%%%%%%%%%%%%%%%%%%%%%%%%%%%

 \bibliography{stretch}

\ifx\mainismaster\UnDef%
    \end{document}
     \fi
%%%%%%%%%%%%%%%%%%%%%%%%%%%%%%%%%%%%%%%%%%